# True-MCSA: A Framework for Truthful Double Multi-Channel Spectrum Auctions

Zhili Chen[1,2], He Huang[3], Liusheng Huang[1,2]
1. School of Computer Science and Technology, University of Science and Technology of China, Hefei, Anhui 230027
2. Suzhou Institute for Advanced Study, University of Science and Technology of China, Suzhou, Jiangsu 215123
3. School of Computer Science and Technology, Soochow University, Suzhou, Jiangsu 215006

*Abstract*—We design a framework for truthful double multi-channel spectrum auctions where each seller (or buyer) can sell (or buy) multiple spectrum channels based on their individual needs. Open, market-based spectrum trading motivates existing spectrum owners (as sellers) to lease their selected idle spectrum channels to new spectrum users (as buyers) who need the spectrum desperately. The most significant requirement is how to make the auction economic-robust (truthful in particular) while enabling spectrum reuse to improve spectrum utilization. Additionally, in practice, both sellers and buyers would require to trade multiple channels at one time, while guaranteeing their individual profitability. Unfortunately, none of the existing designs can meet all these requirements simultaneously. We address these requirements by proposing True-MCSA, a framework for truthful double multi-channel spectrum auctions. True-MCSA takes as input any reusability-driven spectrum allocation algorithm, introduces novel virtual buyer group (VBG) splitting and bidding algorithms, and applies a winner determination and pricing mechanism to achieve truthfulness and other economic properties while improving spectrum utilization and successfully dealing with multi-channel requests from both buyers and sellers. Our results show that the auction efficiency is impacted by the economic factors with efficiency degradations within 30%, under different experimental settings. Furthermore, the experimental results indicate that we can improve the auction efficiency by choosing a proper bidding algorithm and using a base bid. True-MCSA makes an important contribution on enabling spectrum reuse to improve auction efficiency in multi-channel cases.

## I. INTRODUCTION

The demand for radio spectrum use has been growing rapidly with the dramatic development of the mobile telecommunication industry in the last decades. However, the growth of wireless networks has been hampered by the previous inefficient spectrum distribution. In the past decade, the FCC (Federal Communications Commission) and its counterparts across the world have been using single-sided auctions to assign spectrum to wireless service providers in terms of predetermined national/regional long-term leases. The static allocations have led to an artificial shortage of spectrum: new wireless applications starve for spectrum, while large chunks of it remain idle most of the time under their current owners. This allocation inefficiency has prompted a wide interest in an open, marked-based approach for redistributing the spectrum where new users can gain access to the spectrum they desperately need and existing owners can gain financial incentives to "lease" their idle spectrum. Additionally, as the development of multi-radio wireless networks, which is regarded as an enabling technology of next generation wireless network communications, the multi-channel requirement from one single user in the spectrum distribution becomes more and more popular.

Spectrum auctions are among the best-known market-based spectrum allocation mechanisms due to their perceived fairness and allocation efficiency: everyone has an equal opportunity to win and the spectrum channels are sold to bidders who value them the most. Unlike conventional FCC-style auctions that target long-term national/regional leases to service providers, the spectrum auctions we addressed in this paper allow sellers to lease their idle spectrum channels to buyers that can be small wireless networks, individual infrastructure networks or home networks, and provide a promising solution for efficient dynamic spectrum redistributions. There have been some researches [1-4] for the spectrum auctions targeting spectrum redistribution. However, none of them provides a truthful double multi-channel spectrum auction solution.

In this paper, we propose a framework for TRUthful doublE Multi-Channel Spectrum Auctions (True-MCSA) where each seller or buyer requests arbitrary number of spectrum channels to sell or buy based on their individual needs. True-MCSA takes as input any reusability-driven spectrum allocation algorithm, introduces novel virtual buyer group (VBG) splitting and bidding algorithms, and applies a novel winner determination and pricing mechanism to achieve truthfulness and other economic properties while improving spectrum utilization and successfully dealing with multi-channel requests from both buyers and sellers. True-MCSA provides a simple framework to address truthful double multi-channel spectrum auctions with spectrum reuse.

The paper makes the following key contributions.

(1) We propose a framework, True-MCSA, for truthful double multi-channel spectrum auctions. In True-MCSA, we successfully deal with multi-channel requests from both sellers and buyers by introducing virtual buyer group (VBG) splitting and bidding together with applying a novel winner determination and pricing mechanism. True-MCSA provides an efficient and trust-worthy environment for spectrum sellers and buyers to trade arbitrary number of spectrum channels. Through the auctions, each seller sells all the channels it bid if winning, sells none if losing; each buyer buys at most the number of channels it bid if winning, buy nothing if losing.

(2) We formally prove that True-MCSA is of three key economic-properties, namely individual rationality, ex-post

budget balance and truthfulness for both sellers and buyers. These properties ensure that the auctions are economic-robust, the auctioneer have incentives to setup an auction while the bidders have incentives to bid in the auction.

(3) We do extensive experiments to show the auction efficiency compared to that of the pure allocation (PA) algorithm, and the impacts of different bidding algorithms, different bidding patterns and different buyer distributions on the auction efficiency.

The rest of paper is organized as follows: Section II is the preliminaries; Section III describes the concept and design of Ture-MCSA; in section IV, we carry out extensive simulation experiments to evaluate the performance of our design; Section V introduces the related works; finally, Section VI is the conclusions and future works.

## II. PRELIMINARIES

In this section, we provide the problem model of the double multi-channel spectrum auction, and discuss our design goals to implement the auction.

### A. Problem Model

We consider a single-round double multi-channel spectrum auction with one auctioneer, $M$ sellers, and $N$ buyers. We assume that each seller contributes multiple distinct channels and each buyer requests multiple channels. The auction is sealed-bid and private. Bidders submit their bids privately to the auctioneer without any knowledge of others.

For a seller $m$, its bid is denoted by $(s_m, c_m)$ ($s_m > 0$ and $c_m \geq 1$), meaning that $m$ require the minimum per-channel payment $s_m$ to sell $c_m$ channels; $v_m^s$ and $c_m^t$ are its true valuation of each channel and true number of channels provided; $p_m^s$ is the per-channel payment received if it wins the auction; and its utility is $u_m^s = c_m^w \cdot (p_m^s - v_m^s)$ if it wins $c_m^w$ ($1 \leq c_m^w \leq c_m$) channels in the auction, and 0 otherwise. For a buyer $n$, its bid is denoted by $(b_n, d_n)$ ($b_n > 0$ and $d_n \geq 1$), which represents that the buyer is willing to pay the maximum price $b_n$ for each channel, and it requires $d_n$ channels; $v_n^b$ and $d_n^t$ are its true valuation of each channel and true number of channels requested; $p_n^b$ is the per-channel price it pays if it wins the auction; and its utility is $u_n^b = d_n^w \cdot (p_n^b - v_n^b)$ if it wins $d_n^w$ ($1 \leq d_n^w \leq d_n$) channels in the auction, and 0 otherwise.

Note that in the auctions, we assume that both sellers and buyers can bid the per-channel price untruthfully, while the buyers can also bid the number of channels requested untruthfully and the sellers always bid the true number of channels he can provide. We assume that when $d_n^w > d_n^t$, the utility of each extra channel for buyers is not more than zero.

### B. Design Goals

Our first design goal is to exploit the spatial reusability of radio spectrum. Unlike conventional goods, spectrum is reusable among bidders subjecting to the spatial interference constraints: bidders in close proximity cannot use the same spectrum frequency simultaneously but well-separated bidders can. In the case of multi-channel spectrum auctions, different buyer request quite different number of spectrum channels, how to exploit and maximize the spatial reusability is challenging.

Our second design goal is to ensure economic-robustness of the auctions. Truthfulness, individual rationality and budget balance are the three critical properties required to design economic-robust double auctions [1][5][6]. Although TRUST proposed by [1] has well achieved all the three properties in one-channel spectrum auctions, how to achieve these economic properties in multi-channel spectrum auctions has not been addressed. In the multi-channel spectrum auctions, both sellers and buyers request different numbers of channels, which makes it challenging to design the auction process (i.e. determination of winners and prices) and to achieve economic robustness.

We now define the three economic properties in double multi-channel spectrum auctions:

(1) Truthfulness. A double multi-channel spectrum auction is truthful if no matter how other players bid, no seller $m$ or buyer $n$ can improve its own utility by biding untruthfully ($s_m \neq v_m^s$ for sellers and $b_n \neq v_n^b$ or $d_n \neq d_n^t$ or both for buyers).

Truthfulness is essential to avoid market manipulation and ensure auction fairness and efficiency. In untruthful auctions, selfish bidders can manipulate their bids to game the system to increase their utilities but decrease others'. In truthful auctions, the dominate strategy for bidders is to bid truthfully, thereby eliminating the fear of market manipulation and the overhead of strategizing over others. With the true valuations, the auctioneer can allocate spectrum efficiently to buyers who value it the most.

(2) Individual Rationality. A double multi-channel spectrum auction is individual rational if no winning seller is paid less than its bid and no winning buyer pays more than its bid:

$$p_m^s \cdot c_n^w \geq s_m \cdot c_n^w, \quad p_n^b \cdot d_n^w \leq b_n \cdot d_n^w \qquad (1)$$

Here, we assume the pricing is uniform for each seller $m$ and buyer $n$, which is in accordance with our design.

This property guarantees non-negative utilities for bidders who bid truthfully, providing them incentives to participate.

(3) Ex-post Budget Balance. A double multi-channel spectrum auction is ex-post budget balanced if the auctioneer's profit $\Phi \geq 0$. The profit is defined as the difference between the revenue collected from buyers and the expense paid to sellers:

$$\Phi = \sum_{n=1}^{N} p_n^b \cdot d_n^w - \sum_{m=1}^{M} p_m^s \cdot c_n^w \geq 0 \qquad (2)$$

This property ensures that the auctioneer has incentives to set up the auction.

### III. TRUE-MCSA: CONCEPT AND DESIGN

In this section, we first describe the concept of designing True-MCSA, then present the design in detail, finally, prove the auction properties that our design satisfies.

#### A. Concept

The most challenging problem in designing True-MCSA is how to deal with the multi-channel requirements of both buyers and sellers, while guaranteeing that the double spectrum auctions are truthful, and the reuse of spectrum is well exploited. We borrow ideas from McAfee's design [7] and TRUST [1], and propose a novel auction framework that meets all the above requirements. Specifically, we form buyer groups independently on buyer bids to exploit the reuse of spectrum; design VBG splitting and VBG bidding algorithms to solve the problem of multi-channel bidding; bring forward a novel winner determination mechanism to ensure truthful double spectrum auctions.

**(1) Bid-independent Buyer Grouping**

The first question is how to group multiple conflict-free buyers together so that they can be assigned the same channels. This spectrum allocation process can be dependent on the bids, like VERITAS [2]. However, a bid-dependent allocation allows bid manipulation and makes the auctions untruthful [1]. Therefore, we take the same policy as TRUST, and form the buyer groups based on their interference conditions but independent of their bids. The buyer grouping initially exploits the special reuse of spectrum among buyers located in different places.

**(2) Virtual Buyer Group (VBG) Splitting and Bidding**

After forming buyer groups, we can not directly treat each buyer group as a super buyer like TRUST, for in multi-channel scenarios each buyer in the group may request quite different number of channels and it is hard to determine the group bid and how many channels the buyer group should buy. Our basic idea is, for each buyer group, we should first properly split it into several VBGs in which each buyer merely request one channel, and then regard each VBG as a super buyer to bid.

Based on the basic idea, we form VBGs from a buyer group like this: the first VBG is obtained by gathering buyers requesting the first channels in the buyer group; the second one is obtained by gathering buyers requesting the second channels in the buyer group; ...and son on. In this way, a buyer group is split into $K$ VBGs, where $K$ is the maximum number of channels requested by the buyers in the buyer group. Finally, a VBG is treated as a super buyer and the problem settings are reduced to those in McAfee's design.

Though we have split buyer groups into VBGs and reduced the problem settings to simpler ones, it is far from enough to achieve truthfulness. The next question encountered is how to design the bidding for each VBG to participate in the auctions. As discriminatory pricing leads to untruthful auctions [1], uniform pricing should be used to charge buyers in each VBG if it is winning. Additionally, VBGs derived from the same buyer group contains the buyers from the same buyer set. Making use of the above considerations, we design the methods of VBG bidding. In Section III-B, two methods of VBG bidding are proposed.

Through the VBG splitting and bidding, we convert the problem of multi-channel auctions to that of single-channel auctions, and thus properly solve the multi-channel request problems. Furthermore, as we will see in Section III-B, the VBG splitting and bidding also answers the question of how many channels a buyer group should buy while maximizing the spectrum reuse and auction efficiency.

**(3) Winner Determination**

To avoid the bid manipulation and ensure the economic-robustness of the spectrum auctions, the winner determination should lead to the following results: (1) the price charged to the winning buyers is not more than and independent on the per-channel bid of each winning buyer; (2) the price paid to the winning sellers is not less than and independent on the per-channel bid of each winning seller; (3) the pricing mechanisms for both buyers and sellers should be uniform globally or locally. In Section III-B, we provide a novel winner determination that meets the above condition all and properly deals with the multi-channel scenarios.

#### B. Design

Now, we present the design of True-MCSA in details. During the presentation, the following example is used to illustrate the auction process.

**An Example:** In an auction, we assume that the seller set $S$ and the buyer set $B$ with their bids are as follows:

$$S = \{S_1(3,1), S_2(4,2), S_3(5,3), S_4(6,2), S_5(11,2)\}$$
$$B = \{B_1(10,3), B_2(8,5), B_3(5,1), B_4(3,2), B_5(11,2),$$
$$B_6(9,4), B_7(5,1)\}$$

We will discuss how the example auction proceeds in each step of True-MCSA.

True-MCSA consists of the following four steps.

**Step I: Buyer Group Formation**

We assume that all the sellers' channels are available to all the buyers and use conflict graph to describe the interference condition among buyers. Buyers that do not interfere with each other are grouped into the same group and can be assigned to the same channels. The group formation is performed privately by the auctioneer before the actual auction and kept confidential to the buyers. Modeling the interference condition as conflict graph, the group formation is equivalent to finding the independent sets of the conflict graph [8][9]. It is noted that the group formation only forms buyer groups, but not assigns specific channels to buyers.

**Example Illustration**: In this step, we just simply assume that the buyers are grouped into two groups: $G_1 = \{B_1, B_2, B_3, B_4\}$ and $G_2 = \{B_5, B_6, B_7\}$.

**Step II: VBG Splitting and Bidding**

Buyers in groups request to buy multiple channels. We denote the bid of buyer $i$ by $(b_i, d_i)$, where $b_i$ is the per-channel bid and $d_i$ is the number of channels requested. Assuming that buyer $i$ belongs to group $G_n$, the maximal number of channels requested in the group is denoted by $K_n = \max_{i \in G_n} d_i$. Thus, we can split the buyer group $G_n$ into $K_n$ VBGs in which each buyer requests only one channel as follows:

The $1^{st}$ VBG consists of the buyers in group $G_n$ who request their $1^{st}$ channel;

the $2^{nd}$ VGB consists of the buyers in group $G_n$ who request their $2^{nd}$ channel;

…

the $K_n^{th}$ VBG consists of the buyers in group $G_n$ who request their $K_n^{th}$ channel.

Fig.1 illustrates the VBG splitting procedures. Buyer group G = {1, 2, 3, 4} is split into 5 VBGs according to the number of requested channels of each buyer. Since the VBGs with lower indexes are always the super sets of those with higher indexes and thus have higher bids, they have greater probabilities to win in the auction. In other words, buyers in the buyer group tend to win the first several channels while losing the last ones. The reason is that as the increase of the number of channels the buyer group wins, the spectrum reuse decreases in the group, and the auction should find a proper traded channel number for each buyer group by bid competition among VBGs derived from all the buyer groups, maximizing the total spectrum reuse and the auction efficiency.

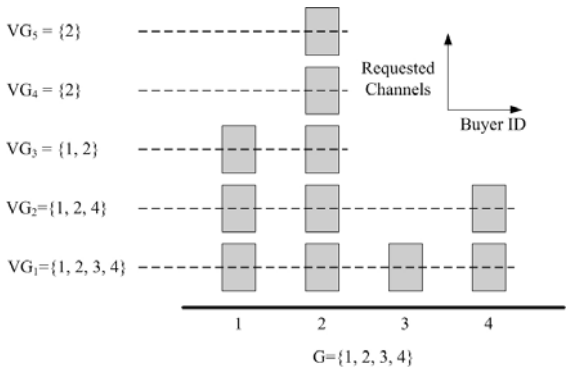

Fig. 1 A Virtual Buyer Group Splitting Illustration.

The VBG splitting equivalently transforms a buyer group to a number of virtual buyer groups, decomposing the complicated multi-channel requesting cases to simple one-channel requesting ones.

After splitting buyer groups into VBGs, we get to design the VBG bidding algorithms. But before this, we should first introduce the notion of critical buyers as follows.

*Definition: A **Critical Buyer** is a buyer whose per-channel bid determines the per-channel price for the buyers in its buyer group.*

From the definition, we can see that a critical buyer determines the bid of each VBG derived from its buyer group. However, the critical buyer of a group is not determined until the design criterion of VBG bidding is chosen.

The notion of critical buyers plays a key role in our design of VBG bidding. With this notion, we design two methods for VBG bidding, namely member-minimized biding and group-maximized bidding. The main difference of the two VBG bidding methods is to choose different critical buyers to meet different design goals.

---

**Algorithm 1 VBG Splitting and MMIN Bidding**

---

1: Function VBGSplitMMBid( $G_n$ )

2:    $b_{\min} = \min_{i \in G_n} b_i$

3:    $i_{\min} = \arg\min_{i \in G_n} b_i$ // critical buyer

4:    $G'_n = G_n - \{i_{\min}\}$

5:    $K_n = \max_{i \in G'_n} d_i$

6:    $G^n = \phi$ // set of VBGs derived from group $G_n$

7:    $\pi^n = \phi$ // set of bids corresponding to $G^n$

8:    for $i = 1 \to K_n$ do

9:      $G_i^n = \phi$ // $i^{th}$ VBG

10:      foreach $j \in G'_n$ do

11:        if $d_j \geq i$ then

12:          $G_i^n = G_i^n \cup \{j\}$

13:        end if

14:      end foreach

15:      $\pi_i^n = b_{\min} \cdot |G_i^n|$

16:      $G^n = G^n \cup \{G_i^n\}$

17:      $\pi^n = \pi^n \cup \{\pi_i^n\}$

18:    end for

---

**1) Member-Minimized (MMIN) Biding**

MMIN bidding targets to maximize the number of buyers selected in each buyer group to participate in the winner determination. In MMIN bidding, the critical buyer in buyer group $G_n$ is identified to be the buyer with minimal per-channel bid in the group (if more than one, one is randomly selected) and the per-channel bid is denoted by $b_{\min}$. The critical buyer is then eliminated from each VBG derived from $G_n$ if existing. Then, the bid of each VBG is calculated by multiplying $b_{\min}$ to the number of its virtual buyers after eliminating.

$$\pi_l = b_{\min} \cdot |G_l^n| \tag{3}$$

Algorithm 1 shows the algorithm of VBG splitting and bidding using MMIN bidding. The algorithm outputs the set $G^n$ of VBGs derived from buyer group $G_n$ and its corresponding bid set $\pi^n$.

**2) Group-Maximized (GMAX) Bidding**

GMAX bidding aims to maximize the first VBG bid by selecting a proper critical buyer. The bid of the first VBG of buyer group $G_n$ is defined as

$$\pi_1 = \max_{i \in G_1'^n, i \geq 2} b_i \cdot (i\text{-}1), \text{ with } I_n = \arg\max_{i \in G_1'^n, i \geq 2} b_i \cdot (i\text{-}1) \tag{4}$$

Where $G_1'^n$ is obtained from the first VBG $G_1^n$ by sorting its buyers in non-increasing order in term of per-channel bid and $i$ is the buyer rank starting from 1. Then buyer $I_n$ in $G_1'^n$ is the critical buyer. The buyers in buyer group $G_n$ with per-channel bids smaller than that of the critical buyer, together with the critical buyer itself are eliminated for their low bids, from each VBG derived from $G_n$. Doing this guarantees that the first VBG bids a maximized bid and the entire VBGs bid independently on the per-channel bid of each buyer left after eliminating. The bids of other VBGs of buyer group $G_n$ can be calculated by multiplying the number of their buyers to the per-channel bid of the critical buyer.

It is obvious that all the buyer groups containing only one buyer will be eliminated from participation in the winner determination.

**Tab.1** The Procedure of VBG Splitting and Bidding

| Group | VBG | Member Set | Selected Set | Bid |
|---|---|---|---|---|
| $G_1$ ($B_4$) | $G_{11}$ | {$B_1$(10), $B_2$(8), $B_3$(5), $B_4$(3)} | { $B_1$(10), $B_2$(8), $B_3$(5)} | 9 |
| | $G_{12}$ | { $B_1$(10), $B_2$(8), $B_4$(3)} | { $B_1$(10), $B_2$(8)} | 6 |
| | $G_{13}$ | { $B_1$(10), $B_2$(8)} | { $B_1$(10), $B_2$(8)} | 6 |
| | $G_{14}$ | { $B_2$(8)} | { $B_2$(8)} | 3 |
| | $G_{15}$ | { $B_2$(8)} | { $B_2$(8)} | 3 |
| $G_2$ ($B_7$) | $G_{21}$ | { $B_5$(11), $B_6$(9), $B_7$(5)} | { $B_5$(11), $B_6$(9)} | 10 |
| | $G_{22}$ | { $B_5$(11), $B_6$(9)} | { $B_5$(11), $B_6$(9)} | 10 |
| | $G_{23}$ | { $B_6$(9)} | { $B_6$(9)} | 5 |
| | $G_{24}$ | { $B_6$(9)} | { $B_6$(9)} | 5 |

**Example Illustration**: In this step, we use MMIN bidding as example. The critical buyers in both $G_1$ and $G_2$ are identified to be $B_4$ and $B_7$, respectively; then the two groups are split into VBGs according to the buyers' requested numbers of channels and critical buyers $B_4$ and $B_7$ are eliminated from these VBGs; finally, the bid of each VGB is calculated by multiplexing the size of selected set to the per-channel bid of its buyer group's critical buyer, as illustrated in Tab. 1.

---

**Algorithm 2 Winner Determination**
---
1: Preconditions:
2: $S = \{s_1, s_2, ..., s_M\}, s.t. \ s_1 \leq s_2 \leq ... \leq s_M$
3: $C = \{c_1, c_2, ..., c_M\}$ is the set of requested number for $S$
4: $\pi = \{\pi_1, \pi_2, ..., \pi_K\}, s.t. \ \pi_1 \geq \pi_2 \geq ... \geq \pi_K$
5: Function WinnerDetermine( $S$, $C$, $\pi$ )
6:     $L = \sum_{i=1}^{M} c_i$
7:     for $i = 1 \rightarrow \min\{L, K\}$ do
8:        $j = 1 + \arg\max_{0 \leq h \leq M} \{\sum_{l=1}^{h} c_l < i\}$
9:        $sum_1 = \sum_{l=1}^{i} \pi_l$
10:      $sum_2 = i \cdot s_j$
11:      if $sum_1 < sum_2$ then
12:        break
13:      end if
14:    end for
15:    $i = i - 1$ // last profitable trade
16:    $j = \arg\max_{0 \leq h \leq M-1} \{\sum_{l=1}^{h} c_l < i\}$ // last seller winner
17:    $k = \sum_{l=1}^{j} c_l$ // last buyer winner

---

**Step III: Winner Determination**

In this step, we determine the winning VBGs that would buy spectrum channels. We assume that spectrum channels are homegenious. In the multi-channel spectrum auction, a seller provides multiple channels and bids a per-channel bid, that is bidding $(s_m, c_m)$. The sellers' per-channel bids are sorted by in non-decreasing order and the buyer (namely VBG in this step) bids are sorted by VBG bid in non-increasing order:

$$\begin{aligned} S &: s_1 \leq s_2 \leq ... \leq s_M \\ \pi &: \pi_1 \geq \pi_2 \geq ... \geq \pi_K \end{aligned} \tag{5}$$

In the case of ties, the ordering is random, with each tied seller or buyer bidder having an equal probability of being ordered prior to the other one.

Let $L = \sum_{i=1}^{M} c_i$ is the total number of channels provides by sellers, $j(i) = 1 + \arg\max_{0 \leq h \leq M-1} \{\sum_{l=1}^{h} c_l < i\}$ is the seller in the $i^{\text{th}}$ trade, the last profitable trade $k'$ is defined as:

$$k' = \arg\max_{i \leq \min\{L, K\}} \{\sum_{l=1}^{i} \pi_i \geq i \cdot s_j\} \tag{6}$$

Then the auction winners are the first $k = \sum_{l=1}^{j(k')-1} c_l \leq k'-1$ VBGs in $\pi$ and the first $j(k)$ ($= j(k')-1$) sellers in $S$. The winning sellers lease channels to the winning VBGs, one for each one. For each winning buyer, the number of channels he buys is the number of winning VBGs to which he belongs to; for each winning seller, the number of channels he sells is always the number of channels he bids. The winner determining process is described in Algorithm 2.

**Example Illustration**: Tab. 2 shows the procedure of winner determination. In order to make presentation simple, we rewrite each seller as many times as the number of channels it bid and sort the sellers in non-decreasing order in term of per-channel bid. Then, all the VBGs are sorted in non-increasing order in term of VBG bid. According to the bid accumulation (bid acc.), at the place of ($S_4$, $G_{14}$), Equation (6) is satisfied. Thus, we get the last trade $k'=8$, the last winning seller $j=3$, and the last winning VBG $k=1+2+3=6$. So the number of winning VBGs is 6, and the number of winning sellers is 3. The shaded cells in the table indicate that the last two profitable trades are sacrificed for truthfulness. The results (WS: Winning Sellers, WVBG: Winning Virtual Buyer Groups, WB: Winning Buyers) of winner determination are summarized in the last three rows of Tab. 1, where the winning channel number and the requesting channel number are separated by "/", e.g. $B_2(8,3/5)$ means that $B_2$ bids a per-channel bid 8 and win 3 channels out of 5 requesting channels.

**Tab. 2** The Procedure of Winner Determination

| NO. | 1 | 2 | 3 | 4 | 5 | 6 | 7 | 8 | 9 | 10 |
|---|---|---|---|---|---|---|---|---|---|---|
| Sellers | $S_1$ | $S_2$ | $S_2$ | $S_3$ | $S_3$ | $S_3$ | $S_4$ | $S_4$ | $S_5$ | $S_5$ |
| Bids | 3 | 4 | 4 | 5 | 5 | 5 | 6 | 6 | 11 | 11 |
| Bid-Acc. | 3 | 8 | 12 | 20 | 25 | 30 | 42 | 48 | 99 | 110 |
| VBGs | $G_{21}$ | $G_{22}$ | $G_{11}$ | $G_{12}$ | $G_{13}$ | $G_{23}$ | $G_{24}$ | $G_{14}$ | $G_{15}$ | - |
| Bids | 10 | 10 | 9 | 6 | 6 | 5 | 5 | 3 | 3 | - |
| Bid-Acc. | 10 | 20 | 29 | 35 | 41 | 46 | 51 | 54 | 57 | - |
| Res WS | $S_1, S_2, S_3$ | | | | | | | | | |
| Res WVBG | $G_{21}, G_{22}, G_{11}, G_{12}, G_{13}, G_{23}$ | | | | | | | | | |
| Res WB | $B_1(10,3/3), B_2(8,3/5), B_3(5, 1/1), B_5(11, 2/2), B_6(9, 3/4)$ | | | | | | | | | |

**Step IV: Pricing**

Each buyer in the same winning VBG is charged an even share of the VBG bid and each channel is paid by the price $s_{j(k)}$. Then, each buyer is charged the sum of what it is charged in all the winning VBGs it belongs to and each seller is paid by the product of multiplying the number of winning channels he bid to the price $s_{j(k)}$.

**Tab. 3** The Utility Calculation of Each Seller and Buyer

| Seller | $S_1$ | $S_2$ | $S_3$ | - | - |
|---|---|---|---|---|---|
| Pay. | 6×1=6 | 6×2=12 | 6×3=18 | - | - |
| Util. | 6-3=3 | (6-4)×2=4 | (6-5)×3=3 | - | - |
| Buyer | $B_1$ | $B_2$ | $B_3$ | $B_5$ | $B_6$ |
| Charg. | 3×3=9 | 3×3=9 | 3×1=3 | 5×2=10 | 5×3=15 |
| Util. | (10-3)×3=21 | (8-3)×3=15 | 5-3=2 | (11-5)×2=12 | (9-5)×3=12 |

**Example Illustration**: Tab. 3 lists the calculation of utility for each seller and buyer. It is obvious that the utility of each seller and buyer is positive, so the individual rationality is satisfied. From Tab. 2, we can see that ex-post budget balance is also satisfied.

### C. Proof of Auction Properties

In this section, we prove that True-MCSA satisfies the properties of ex-post budget balance, individual rationality and truthfulness for both sellers and buyers. We only prove the case when using MMIN VBG bidding, while the proof of the case when using GMAX VBG bidding is similar and we omit it for limitation of space.

**1) Proof of Ex-post Budget Balance:**

**Theorem 1**: True-MCSA is ex-post budget balanced i.e. $\Phi \geq 0$.

**Proof**: Because $k$ is the number of winning VBGs and the number of the channels traded in the auctions, and $k$ satisfies $\sum_{i=1}^{k} \pi_i \geq k \cdot s_{j(k)}$, thus $\Phi = \sum_{i=1}^{k} \pi_i - k \cdot s_{j(k)} \geq 0$. □

**2) Proof of Individual Rationality:**

**Theorem 2:** True-MCSA is individual rational.

**Proof**: By the definition of individual rationality, we need to show that no winning seller will be paid less than its bid, and no winning buyer will be charged more than its bid.

First, because True-MCSA sorts seller's per-channel bids in a non-decreasing order and pays each winning seller $m$ with last profitable seller $j(k)$'s per-channel bid, the payment to $m$ is $p_m^s \cdot c_m^w = s_{j(k)} \cdot c_m^w \geq s_m \cdot c_m^w$, where $c_m^w$ is the number of channels that seller $m$ manages to sell and it always satisfies $c_m^w = c_m$ in True-MCSA. Second, for each winning buyer $n$, the per-channel price charged to $n$ is $p_n^b = b_{\min} \leq b_n$, where $b_{\min}$ is the smallest buyer per-channel bid in $n$'s buyer group. Then, the price charged to buyer $n$ is $p_n^b \cdot d_n^w \leq b_n \cdot d_n^w$, where $d_n^w$ is the number of channels that buyer $n$ wins. □

**3) Proof of Truthfulness:**

To prove True-MCSA's truthfulness, we need to show that for any buyer $n$ or seller $m$, it can not improve its utility by bidding other than its true valuation. For this, we first show that its winner determination is monotonic for both sellers and buyers and the pricing is bid-independent. Using these two claims, we then prove the truthfulness.

**(1) Monotonic winner determination**

The following two lemmas summarize the monotonicity of True-MCSA's winner determination.

**Lemma 1**: Given $\{(s_m, c_m)\}_{m=1}^{M}$ and $\{(b_1, d_1),...,(b_{n-1}, d_{n-1}), (b_{n+1}, d_{n+1}),...,(b_N, d_N)\}$, if buyer $n$ wins $d_n^w$ ($0 < d_n^w < d_n$) channels by bidding $(b_n, d_n)$, then, by bidding $(b_n', d_n')$ with $b_n' \geq b_n$ and $d_n' \geq d_n$, buyer $n$ also wins the same number of channels.

**Proof**: Since buyer $n$ wins $d_n^w$ channels by bidding $(b_n, d_n)$, it is not eliminated from its buyer group and $d_n^w$ of its VBGs win the auction. When bidding $(b_n', d_n')$ with $b_n' \geq b_n$ and $d_n' \geq d_n$, buyer $n$ is still not eliminated from its buyer group and the bids of its first $d_n$ VBGs remain the same as before, while the bids of its last $(d_n' - d_n)$ VBGs must be not greater than that of its $d_n^{\text{th}}$ VBG. So, it must be that the same $d_n^w$ ($0 < d_n^w < d_n$) of its VBGs win the auction. Lemma 1 holds. □

**Lemma 2**: Given $\{(b_n, d_n)\}_{n=1}^N$ and $\{(s_1, c_1), ..., (s_{m-1}, c_{m-1}), (s_{m+1}, c_{m+1}), ..., (s_M, c_M)\}$, if seller $m$ wins the auction by bidding $(s_m, c_m)$, then, by bidding $(s_m', c_m)$ with $s_m' < s_m$, seller $m$ also wins the auction.

**Proof**: Since sellers are ranked in non-decreasing order in term of per-channel bid, seller $m$ wins the auction by bidding $(s_m, c_m)$, it must also win by bidding $(s_m', c_m)$ with $s_m' < s_m$. □

**(2) Bid-independent pricing**

We show that the pricing is bid-independent for both winning buyers and sellers.

**Lemma 3**: Given $\{(s_m, c_m)\}_{m=1}^M$ and $\{(b_1, d_1), ..., (b_{n-1}, d_{n-1}), (b_{n+1}, d_{n+1}), ..., (b_N, d_N)\}$, if buyer $n$ wins the same number $d_n^w$ ($0 < d_n^w \leq d_n$) of channels by bidding $(b_n, d_n)$ and $(b_n', d_n')$, then the utility $u_n^b$ for $n$ is the same for both.

**Proof**: It is easy to show that the bids of all the winning VGBs of buyer $n$ remain the same in both cases, then the utility $u_n^b$ for $n$ is the same for both. □

**Lemma 4**: Given $\{(b_n, d_n)\}_{n=1}^N$ and $\{(s_1, c_1), ..., (s_{m-1}, c_{m-1}), (s_{m+1}, c_{m+1}), ..., (s_M, c_M)\}$, if seller $m$ wins the auction by bidding $(s_m, c_m)$ and $(s_m', c_m)$, then the payment $p_m^s$ to $m$ is the same for both.

**Proof**: The proof is similar to that of Lemma 3. □

**Tab. 4** Four possible auction results when bidding truthfully and untruthfully, where X means the bidder loses and √ means it wins.

| Case | 1 | 2 | 3 | 4 |
|---|---|---|---|---|
| The bidder lies | X | X | √ | √ |
| The bidder bids truthfully | X | √ | X | √ |

**(3) True-MCSA's truthfulness**

Using the above claims, we now prove the main results on True-MCSA's truthfulness.

**Theorem 3**: True-MCSA is truthful for buyers.

**Proof**: We need to show that any buyer $n$ cannot obtain higher utility by bidding $(b_n', d_n')$, where $b_n' \neq v_n^b$ or $d_n' \neq d_n^t$ or both. Table 4 lists the four possible auction results for one buyer when it bids truthfully and untruthfully. We now examine these cases one by one.

CASE 1: For both bids, buyer $n$ is denied and charged with zero, leading to the same utility of zero.

CASE 2: This happens only if $b_n' < v_n^b$. Theorem 2 ensures a non-negative utility when $n$ bids truthfully and wins the auction. Thus, its utility is no less than that when it bids untruthfully and loses (zero utility).

CASE 3: This happens only if $b_n' > v_n^b$ while the number of channels bid can be $d_n' = d_n^t$ or $d_n' \neq d_n^t$. Let $\pi_{l_1}, ..., \pi_{l_q}$ and $\pi_{l_1}', ..., \pi_{l_q}'$ represent the bids of the $q$ ($0 < q \leq d_n'$) VBGs, whose auction results are changed from losing (i.e. $n$ wins none of the channels) to winning (i.e. $n$ wins $q$ channels), when $n$ bids truthfully and untruthfully. Then this case can be divided into the following two further cases.

(A) Case $d_n' = d_n^t$

In this case, we have $0 < q \leq d_n^t$. Because $n$ changes the auction results of these VBGs from losing to winning by bidding higher than $v_n^b$, $n$ must be eliminated from its VBGs (no matter its VBGs lose or win the auctions) when he bids truthfully, i.e. $\pi_{l_i} = n_i \cdot v_n^b$, $n_i$ is the size of each VBG, for $i = 1, 2, ..., q$. Conversely, when $n$ bids untruthfully, it is easy to show that the untruthful bids for its VBGs, $\pi_{l_i}'$, must satisfy the following condition: $n_i \cdot b_n' \geq \pi_{l_i}' \geq \pi_{l_i} \geq n_i \cdot v_n^b$. Therefore, the utility when $n$ bids $(b_n', d_n')$ is $\sum_{i=1}^{q}(v_n^b - \pi_{l_i}'/n_i) \leq 0$, which is not more than that when $n$ bids truthfully (0). Theorem 3 holds.

(B) Case $d_n' \neq d_n^t$

This case can only affect the value of $q$. When $d_n' < d_n^t$, it is the same as the case of (A) except that $q$ may become smaller, so the utility when $n$ bids $(b_n', d_n')$ is not more than that when he bids truthfully. When $d_n' > d_n^t$, $q$ may become greater and if $0 < q \leq d_n^t$ the result is the same as case (A), and if $q > d_n^t$, i.e. $q = d_n^t + q'$ ($q' > 0$), the utility when $n$ bids $(b_n', d_n')$ is $\sum_{i=1}^{d_n^t}(v_n^b - \pi_{l_i}^{'}/n_i) + u_{n,e}^b(q^{'}) \leq 0$, where $u_{n,e}^b(q^{'})$ is the utility of the extra $q^{'}$ channels which is not greater than 0 according to our problem assumption. So Theorem 3 holds.

CASE 4: The following two further cases must be examined.

(A) Case $d_n^{'} = d_n^t$

According to Lemmas 1 and 3, it is easy to show that buyer $n$ wins the same number of channels and is charged by the same price in both cases, leading to the same utility.

(B) Case $d_n^{'} \neq d_n^t$

This case can only affect the winning number of channels. When $d_n^{'} < d_n^t$, winning number may become smaller, so the utility when $n$ bids $(b_n^{'}, d_n^{'})$ is not more than that when he bids truthfully. When $d_n^{'} > d_n^t$, if $0 < d_n^w < d_n^t$ ($d_n^w$ is the winning number when $n$ bids truthfully), the winning number of channels must remain the same when $n$ lies according to Lemmas 1, so the result is the same as case (A); and if $d_n^w = d_n^t$, the winning number of channels when $n$ lies must be $d_n^{'w} = d_n^t + \Delta d \geq d_n^t$ ($\Delta d \geq 0$), the utility is $u_n^b(d_n^t) + u_{n,e}^b(\Delta d) \leq u_n^b(d_n^t) = u_n^b(d_n^w)$, i.e. the utility is not greater than that when $n$ bids truthfully, where $u_n^b(d)$ is the utility when $n$ wins $d$ channels and $u_{n,e}^b(\Delta d)$ is the utility of the $\Delta d$ extra winning channels. Theorem 3 holds.

From the above, we show that no buyer can improve its utility by bidding untruthfully. Our proof is completed. □

**Theorem 4:** True-MCSA is truthful for sellers.

**Proof:** Similarly, we need to show that any seller $m$ cannot obtain higher utility by bidding $(s_m^{'}, c_m^{'})$, where $s_m^{'} \neq v_m^s$. Again, the four cases listed in Table 4 are examined.

CASE 1: The same as the buyer case. Theorem 4 holds.

CASE 2: This happens only when $s_m^{'} > v_m^s$. The utility of seller $m$ is non-negative when bids truthfully and wins (Theorem 2), while it is zero when lies and loses. Theorem 4 holds.

CASE 3: This happens only when $s_m^{'} < v_m^s$. First, let $p$ be the per-channel payment to the auction winners when $m$ bids truthfully. Because $m$ loses in this case, $p \leq v_m^s$. Second, let $p^{'}$ be the per-channel payment to the winners (including $m$) when $m$ bids $(s_m^{'}, c_m^{'})$. It is easy to show that because $m$ lowers its bid and wins, $p^{'} \leq p$. Combine the two, we have $p^{'} \leq v_m^s$ and hence $m$'s per-channel utility is $p^{'} - v_m^s \leq 0$ when bidding untruthfully. Thus, no matter the value of $c_m$ is, the utility when $m$ lies is not greater than that when it bids truthfully. Theorem 4 holds.

CASE 4: According to Lemmas 2 and 4, seller $m$ wins the same number of channels and the payment for it does not change, leading to the same utility in both cases. Theorem 4 holds.

Having shown that no seller can improve its utility by bidding other than its true value, our proof completes. □

## IV. EXPERIMENTAL RESULTS

In this section, we use network simulations to evaluate the performance of True-MCSA, and study the auction efficiency of the spectrum auctions and the impacts of bidding algorithms, bidding patterns and buyer distributions on the efficiency.

### A. Simulation Setup

We study the performance of True-MCSA under different settings. The key factors that affect True-MCSA's performance are the underlining spectrum allocation algorithms, the bidding algorithms, bidding patterns, and buyer distributions (i.e. the interference conditions among buyers). For the impact of spectrum allocation algorithms on auction efficiency is similar to that of TRUST, we omit the studies of allocation algorithm and fix it as the algorithm which is based on the maximum independent set of conflict graph. We assume that the buyer interference conditions are modeled by a conflict graph and all buyers are distributed in an area of $100 \times 100$. All the results are averaged over 500 rounds.

#### 1) Bidding algorithms.

In Section III-B, we have proposed two VBG bidding algorithms: MMIN bidding and GMAX bidding. In our experiments, we compare the impacts of the two algorithms on auction efficiency in different settings.

#### 2) Bidding Patterns

We assume that buyers' bids are randomly distributed over $(0,1]$ and those of sellers over $(0,2]$. Also, we assume that auctions can have a base bid value $b_0$ and each bid is then defined by $b = b_0 + \alpha \cdot (b_{\max} - b_0)$, where $\alpha$ is a random number uniformly distributed over $(0,1]$ and $b_{\max} = 1$ for buyers and $b_{\max} = 2$ for sellers. Then, we assume that the channel numbers requested by buyers are randomly distributed over $[1..d_{\max}]$ and those of sellers over $[1..c_{\max}]$, where $[1..X]$ represents the integer number set from 1 to $X$. We use a triple $(c_{\max}, d_{\max}, b_0)$ to represent a bidding pattern for the auctions and examine the impacts of different bidding patterns on the auction efficiency.

#### 3) Buyer Distributions.

The auction efficiency depends on the interference conditions among buyers. We model the interference conditions using a conflict graph, and apply a distance-based criterion to determine whether two buyers conflict. In this case,

the interference condition depends mainly on the buyer distribution. We consider two types of distributions:

**Random Distribution**: We randomly distribute a set of buyers in a given area, with a variety of conflict degrees.

**Clustered Distribution**: We randomly place some buyers in a given area and gradually add buyers in some small center areas, creating some hotspots.

The performance metrics are auction efficiency (or social welfare, $\alpha$), the number of channels traded ($N_t$), the per-channel spectrum efficiency ($\beta$), and the efficiency degradation ($\eta$) over Pure Allocation (PA). Here, auction efficiency is defined as the bid-weighted sum of all the channels won by buyers minus that of all the channels won by sellers as Equation (7). Auction efficiency is in fact the value created by the auction and shared by sellers, buyers and the auctioneer, and thus has another name called social welfare. Since it reflects not only the spectrum reuse but also the financial efficiency of the auctions, we use it as the main performance metric instead of spectrum utilization. What is more, according to [1], spectrum utilization and auction efficiency reflect coarsely the same conclusions, thus we omit the spectrum utilization results. The other metrics are defined as Equations from (8) to (10).

$$\alpha = \sum_{n=1}^{N} b_n \cdot d_n^w - \sum_{m=1}^{M} s_m \cdot c_m^w \quad (7)$$

$$N_t = \sum_{m=1}^{M} c_m^w \quad (8)$$

$$\beta = \alpha / N_t \quad (9)$$

$$\eta = 1 - \alpha / \alpha_{PA} \quad (10)$$

Where $d_n^w$ ($0 \le d_n^w \le d_n$) and $c_m^w$ ($0 \le c_m^w \le c_m$) are the numbers of winning channels of buyer $n$ and seller $m$, respectively, and $d_n^w = 0$ or $c_m^w = 0$ means buyer $n$ or seller $m$ loses; $\alpha_{PA}$ is the auction efficiency of PA.

By default, the experimental settings are as follows: True-MCSA uses MMIN bidding algorithm, random buyer distribution, bidding pattern $(3, 5, 0)$, and the numbers of sellers and buyers are 10 and 100, the protection distance of buyers is 10.

*B. Auction Efficiency*

We start from studying the auction efficiency of the truthful double multi-channel spectrum auctions in different settings. We use PA as a benchmark and change the number of sellers, the number of buyers, and the protection distance of buyers (i.e. the maximal distance that a buyer's radio signal can reach) to evaluate the four performance metrics mentioned above. PA repeatedly selects the unassigned VBG of the biggest size and assigns it a channel regardless of the buyers' bids. For a fair comparison, we implement PA using the same allocation algorithm assuming the number of channels available is equal to the number of channels traded in True-MCSA.

Fig. 2 shows that as the increasing of the number of sellers (other factors are fixed as default), both the auction efficiency and the number of channels traded increase while the per-channel efficiency and efficiency degradation decrease. Since increasing the number of sellers means raising the supplies, we can concludes that raising the supplies can increase the auction efficiency of True-MCSA though decreasing the per-channel efficiency, and the efficiency degradation caused by achieving economic-robustness is diminished, too. Furthermore, the efficiency degradation values are within 30% when the number of sellers changes from 10 to 100.

Fig. 3 illustrates that as the increasing of the number of buyers, the entire performance metrics trend to increase while the degradation over PA has a small-scope fluctuation in its curve. This indicates that raising the demands can increase both the per-channel efficiency and number of channels traded, and thus the auction efficiency, but the efficiency degradation caused by achieving economic-robustness is also increase. Still, the degradation values are within 30% when the number of buyers changes from 20 to 300.

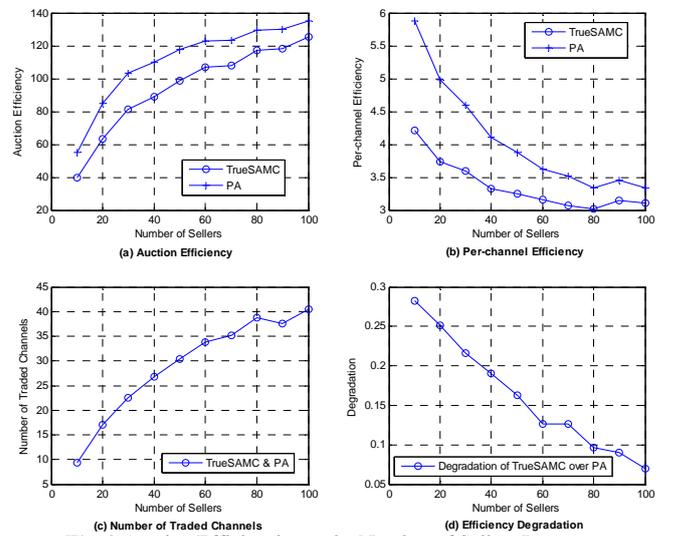
**Fig. 2** Auction Efficiencies as the Number of Sellers Increases

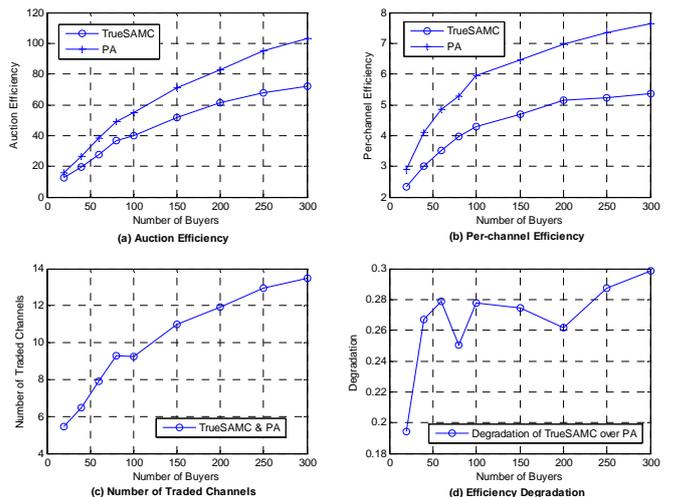
**Fig. 3** Auction Efficiencies as the Number of Buyers Increases

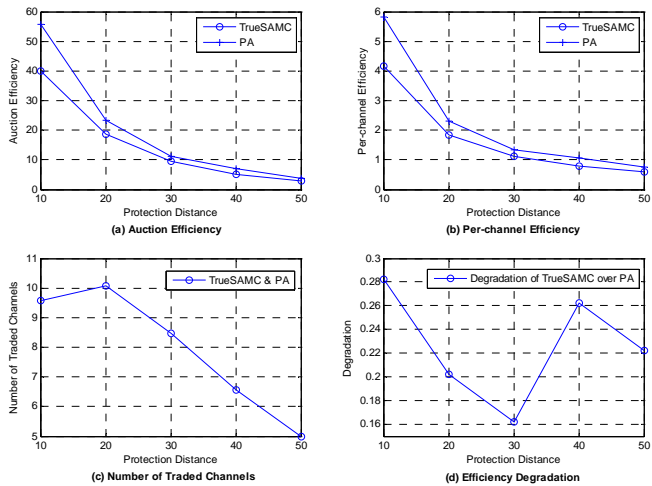

**Fig. 4** Auction Efficiencies as Protection Distance Increases

In Fig. 4, we can see that as the protection distance of buyers increase, the per-channel efficiency and number of channels traded decrease, and thus the auction efficiency decreases, but the efficiency degradation over PA fluctuates within 30%. Therefore, heavy interferences among buyers severely affect the auction efficiency of both Ture-MCSA and PA.

### C. Impact of Bidding Algorithms

In this part, we compare the performances of the two VBG bidding algorithms: MMIN bidding and GMAX bidding.

In Fig. 5, as the increasing of the number of sellers, the per-channel efficiency of MMIN bidding is still greater than that of GMAX bidding, and the traded channel number of MMIN bidding exceeds that of GMAX bidding when the number of sellers is more than 80. As a total effect, the auction efficiency of MMIN bidding exceeds that of GMAX bidding when the number of sellers is more than 30. Furthermore, the efficiency degradation of MMIN bidding decreases more rapidly than that of GMAX bidding and the degradation values of the former are still quite smaller than that of the latter. Therefore, we can conclude that MMIN bidding algorithm is more suitable for auctions with more supplies than GMAX bidding algorithm.

In Fig. 6, it is shown that as the increasing of the number of buyers, the per-channel efficiencies of both MMIN bidding and GMAX bidding increase at the same time, while the former is still greater than the latter; the traded channel numbers of both MMIN bidding and GMAX bidding also increase simultaneously, but the latter is still far greater than the former. As a total effect, the auction efficiency of GMAX bidding is still greater than that of MMIN bidding. Additionally, the efficiency degradation of GMAX bidding becomes smaller than that of MMIN bidding when the number of buyers is greater than 150. Thus, the conclusion is that GMAX bidding algorithm is more suitable for auctions with more demands than MMIN bidding algorithm.

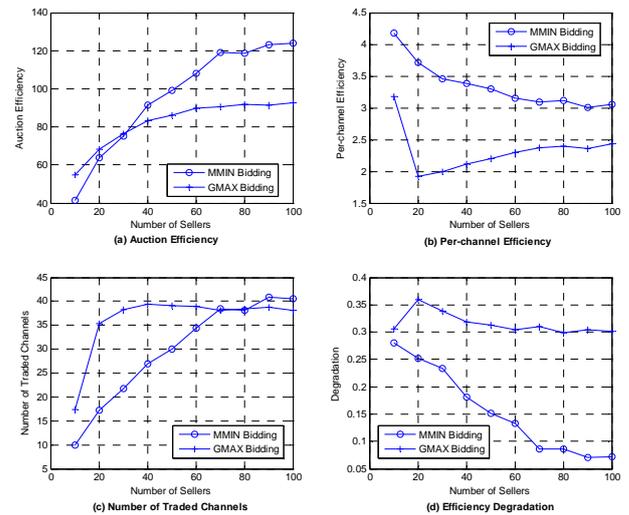

**Fig. 5** Auction Efficiency Comparison as the Number of Sellers Increases

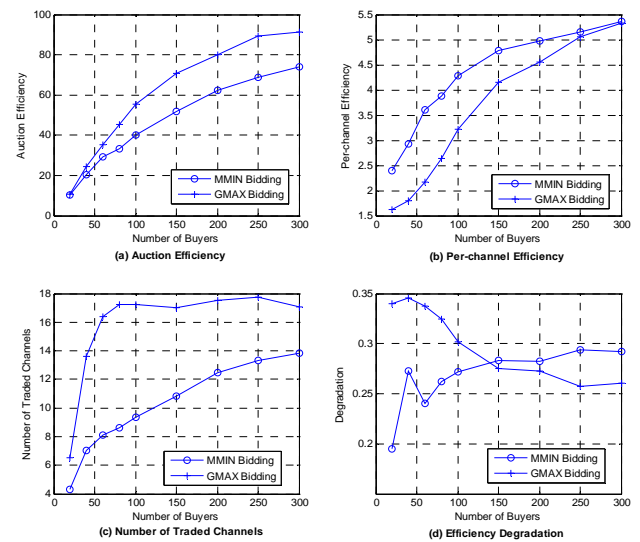

**Fig. 6** Auction Efficiency Comparison as the Number of Buyers Increases

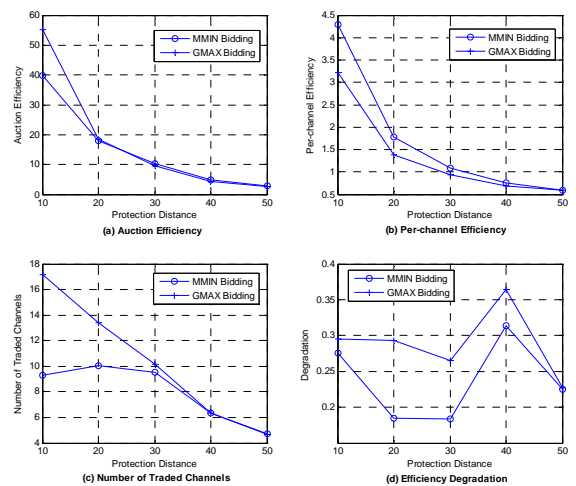

**Fig. 7** Auction Efficiency Comparison as Protection Distance Increases

Fig. 7 shows that as the protection distance increases, the auction efficiency, the per-channel efficiency, and the number of channels traded of both MMIN bidding and GMAX bidding decrease similarly, while the efficiency degradations of both fluctuate similarly, too. Furthermore, all the performance metrics converge to one point as the protection distance is 50. So it can be concluded that the impacts of protection distance for both two bidding algorithms are analogy.

### D. Impact of Bidding Patterns

We also study the impact of different bidding patterns $(c_{\max}, d_{\max}, b_0)$. Tab. 5 lists the performance metrics when applying different bidding patterns. From the table, we can see that: (1) using a positive base bid value like $b_0 = 0.1$ can greatly improve the all the performance metrics; (2) Increasing the value of $c_{\max}$ leads to decreasing of per-channel efficiency but increasing of traded channel number, and thus improving the auction efficiency while slightly decreasing the efficiency degradation; (3) Increasing the value of $d_{\max}$ leads to increasing of both the per-channel efficiency and the traded channel number, and thus improving the auction efficiency while slightly affecting the efficiency degradation. While the values of $c_{\max}$ and $d_{\max}$ depend on the relation between market supply and demand, the auctioneer can greatly improve the auction efficiency and diminish the efficiency degradation by properly designing the base bid value $b_0$.

**Tab. 5** Impact of Bidding Patterns on Auction Efficiency

| NO | Bid Patt. ($c_{max}, d_{max}, b_0$) | Auc. Eff. (α) | Per-Ch. Eff. (β) | Num. Ch. Tr. ($N_t$) | Degrad. (η) |
|---|---|---|---|---|---|
| 1 | (3, 5, 0) | 41.4462 | 4.1571 | 9.9700 | 0.2703 |
| 2 | (3, 5, 0.1) | 75.5006 | 5.0100 | 15.0700 | 0.0773 |
| 3 | (3, 7, 0) | 47.1174 | 4.3668 | 10.7900 | 0.2650 |
| 4 | (3, 7, 0.1) | 80.9589 | 5.2742 | 15.3500 | 0.0858 |
| 5 | (5, 5, 0) | 53.7958 | 3.9153 | 13.7400 | 0.2660 |
| 6 | (5, 5, 0.1) | 94.4755 | 4.6909 | 20.1400 | 0.0564 |
| 7 | (5, 7, 0) | 57.5567 | 3.9234 | 14.6700 | 0.2970 |
| 8 | (5, 7, 0.1) | 102.7039 | 4.8930 | 20.9900 | 0.0740 |

### E. Impact of Buyer Distribution

We randomly place 60 buyers in a given area of size $100 \times 100$ and gradually add 20 buyers in two randomly selected small center areas of size $20 \times 20$, creating 2 hotspots. Tab. 6 shows that the performance metrics suffer declines in different degrees when using cluster distribution.

**Tab. 6** Impact of Buyer Distributions on Auction Efficiency

| Buyer Distr. | Auc. Eff. (α) | Per-Ch. Eff. (β) | Num. Ch. Tr. ($N_t$) | Degrad. (η) |
|---|---|---|---|---|
| Uniform | 40.2154 | 4.2328 | 9.5010 | 0.2743 |
| Cluster | 27.6391 | 3.0223 | 9.1450 | 0.3761 |

## V. RELATED WORKS

Auctions have been widely used to allocate spectrum [10], including transmit power auctions [11], spectrum band auctions [12-15], and spectrum pricing [16-19]. However, these schemes do not consider truthfulness. Paper [2] proposed the first truthful spectrum auction design VERITAS, but only addressed single-sided multi-channel auctions and a direct extension of VERITAS to double auctions is untruthful [1]. For double spectrum auctions, [20] proposed a hierarchical design based on McAfee's design without spectrum reuse. Paper [1] proposed TRUST, the first truthful double auction design with spectrum reuse for multi-party spectrum trading, but TRUST only dealt with one-channel requirements for both sellers and buyers. Paper [3] improved TRUST by redesigning the group bidding and winner determining mechanisms and still only dealt with one-channel requirements as TRUST. Although paper [4] brought forward a simple illustration for solving buyers' two-channel requirements, the solution is computation-prone and hard to achieve and scale. Furthermore, the solution seems to be truthful only for buyers but not for sellers, since the pricing among sellers is impossible to be uniform according to the paper and sellers can manipulate their bids to obtain higher utilities. True-MCSA provides a simple truthful double auction design with spectrum reuse for multi-party multi-channel spectrum trading. In addition, True-MCSA can work with various spectrum allocation algorithms [8], [9], [21].

Truthfulness is a critical factor to attract participation [6]. Many truthful mechanisms have been developed in conventional double auctions including single-unit [22], [23], [7] and multi-unit double auctions [5], [24]. The majority of these designs follow the idea of McAfee's mechanism [7], using the trade reduction to maintain truthfulness. True-MCSA differs significantly from these conventional designs in that it exploits the spectrum reusability to distribute spectrum efficiently while successfully deals with multi-channel requests from both sellers and buyers to better fit practical requirements.

## VI. CONCLUSIONS AND FUTURE WORKS

We propose Ture-MCSA, a framework for truthful double multi-channel spectrum auctions to support dynamic multi-party multi-channel spectrum trading. Ture-MCSA achieves truthfulness, individual rationality, and ex-post budget balance, the three key economic properties required for economic-robust auctions. Furthermore, Ture-MCSA successfully deals with the multi-channel requests from both sellers and buyers, while enables spectrum reuse to significantly improve auction efficiency. From the design and evaluation of Ture-MCSA, we see that the auction efficiency is impacted by the economic factors leading to certain efficiency degradations. However, the experimental results suggest that we can improve the auction efficiency by choosing a proper bidding algorithm and using a base bid. True-MCSA makes an important contribution on enabling spectrum reuse to improve auction efficiency in multi-channel cases.

In this paper, we assume that all the spectrum channels are homogeneous and design True-MCSA to achieve multi-channel spectrum auctions. Through the auctions, sellers sell all the channels they provide when winning but sell nothing when losing, buyers buy at most all the channels they require when winning while buy nothing when losing. However, there are other request formats [2], e.g. requesting continuous channels, requesting restrict number of channels, which should be addressed in future work. Finally, how to resist collusions and what is the tradeoff between auction efficiency and collusion-resisting economic robustness in double truthful multi-channel spectrum auctions are also interesting problems worth exploring in the future.

REFERENCES

[1] X. Zhou and H. Zheng. Trust: A general framework for truthful double spectrum auctions. Proc. of Infocom 2009, pp. 999-1007.

[2] X. Zhou, S. Gandhi, S. Suri, and H. Zheng. eBay in the sky: Strategy-proof wireless spectrum auctions. Proc. of MobiCom 2008, pp.2-13.

[3] E. Yao, L. Lu, W. Jiang. An efficient truthful double spectrum auction design for dynamic spectrum access. Proc. of 6th International ICST Conference on Cognitive Radio Oriented Wireless Networks and Communications (CROWNCOM) 2011, pp. 181-185.

[4] F. Wu and N. Vaidya. Small: A strategy-proof mechanism for radio spectrum allocation. Proc. of Infocom 2011, pp. 81-85.

[5] M. Babaioff, and W.E. Walsh. Incentive-compatible, budget balanced, yet highly efficient auctions for supply chain formation. In Proc. of Economic Commerce (2003).

[6] P. Klemperer. What really matters in auction design. Journal of Economic Perspectives, 16(1): 169–189, Winter 2002.

[7] R.P. Mcafee. A dominant strategy double auction. Journal of Economic Theory, 56(2): 434–450, April 1992.

[8] S. Ramanathan. A unified framework and algorithm for channel assignment in wireless networks. Wireless Networks, 5(2): 81–94, 1999.

[9] A.P. Subramanian, H. Gupta, S.R. Das, and M.M. Buddhikot. Fast spectrum allocation in coordinated dynamic spectrum access based cellular networks. In Proc. of IEEE DySPAN (November 2007).

[10] P. CRAMTON. Spectrum auctions. Handbook of Telecommunications Economics, pp.605–639, 2002.

[11] J. Huang, R. Berry, and M. Honig. Auction mechanisms for distributed spectrum sharing. In Proc. of 42nd Allerton Conference (2004).

[12] M. Buddhikot, And K. Ryan. Spectrum management in coordinated dynamic spectrum access based cellular networks. In Proc. of IEEE DySPAN (2005).

[13] S. Gandhi, C. Buragohain, L. Cao, H. Zheng, and S. Suri. A general framework for wireless spectrum auctions. In Proc. of IEEE DySPAN (2007).

[14] K. Ryan, E. Aravantinos, and M. Buddhikot. A new pricing model for next generation spectrum access. In Proc. of TAPAS (2006).

[15] S. Sengupta, M. Chatterjee, and S. Ganguly. An economic framework for spectrum allocation and service pricing with competitive wireless service providers. In Proc. of IEEE DySPAN (November 2007).

[16] A.A. Daoud, M. Alanyali, and D. Starobinski. Secondary pricing of spectrum in cellular cdma networks. In Proc. of IEEE DySPAN (November 2007).

[17] O. Ileri, D. Samardzija, and N.B. Mandayam. Demand responsive pricing and competitive spectrum allocation via a spectrum server. In Proc. of IEEE DySPAN (2005).

[18] H. Mutlu, M. Alanyali, and D. Starobinski. Spot pricing of secondary spectrum usage in wireless cellular networks. In Proc. of INFOCOM (April 2008).

[19] Y. Xing, R. Chandramouli, And C. Cordeiro. Price dynamics in competitive agile spectrum access markets. IEEE Journal on Selected Areas in Communications, 25(3): 613–621, April 2007.

[20] T. Yamada, D. Burgkhardt, I. Cosovic, And F.K. Jondral. Resource distribution approaches in spectrum sharing systems. EURASIP Journal on Wireless Communications and Networking (2008).

[21] F. Wu, S. Zhong, And C. Qiao. Globally optimal channel assignment for non-cooperative wireless networks. In Proc. of INFOCOM (April 2008).

[22] M. Babaioff, and N. Nisan. Concurrent auctions across the supply chain. In Proc. of Economic Commerce (2001).

[23] K. Deshmukh, A.V. Goldberg, J.D. Hartline, And A.R. Karlin. Truthful and competitive double auctions. In Proc. of ESA (2002).

[24] P. Huang, A. Scheller-Wolf, And K. Sycara. Design of a multi-unit double auction e-market. Computational Intelligence 18, 4 (2002).